# Comment on "Observation of the $\nu = 1$ quantum Hall effect in a strongly localized two-dimensional system"

S. V. Kravchenko

Physics Department, City College of the City University of New York, New York, New York 10031

(April 11, 1996)

In a recent Rapid Communication "Observation of the $\nu = 1$ quantum Hall effect in a strongly localized two-dimensional system," Shahar, Tsui, and Cunningham reported a disorder—magnetic field phase diagram for the integer quantum Hall effect that appeared to differ qualitatively from theoretical predictions as well as from experimental results obtained by others. In this Comment, I suggest a possible origin of this difference.

PACS numbers: 73.40.Hm, 71.30.+h

In a recent paper [1], Shahar, Tsui, and Cunningham observed a magnetically induced direct transition from an insulating state in zero magnetic field, $B = 0$, to a quantum Hall effect (QHE) state with Hall conductivity $\sigma_{xy} = 1e^2/h$ (I will call this transition "$0 \rightarrow 1$"). This is in contrast with previous reports by a number of experimental groups [2-4] which found a transition from insulator to QHE with $\sigma_{xy} = 2e^2/h$ ("$0 \rightarrow 2$") with no traces of QHE at a Landau filling factor $\nu = 1$. Consequently, the disorder-$B$ phase diagram (equivalent to the disorder-$\nu^{-1}$ phase diagram) for the QHE, presented in Ref. [1] and schematically shown in Fig. 1 (c), is qualitatively different from that obtained in other experiments (schematically shown in Fig. 1 (b)). It also does not agree with the one theoretically predicted by Fogler and Shklovskii [5], which has the shape of Fig. 1 (b) and allows for magnetic-field-induced transitions $0 \rightarrow 2 \rightarrow 0$ but does not permit $0 \rightarrow 1 \rightarrow 0$ transitions. However, Fogler and Shklovskii noted that their phase diagram is not valid for $\nu = 1$.

In this Comment I argue that the phase diagram for the QHE is not unique and depends on the range of electron density (and, hence, magnetic field). For simplicity, I will consider the lowest Landau level only. If the spin splitting is zero, the phase diagram looks like the one shown in Fig. 1 (a) — it is just a two-fold version of the phase diagram suggested by Kivelson, Lee, and Zhang for spinless electrons [6]. As the magnetic field increases, one can only observe $0 \rightarrow 2 \rightarrow 0$ transitions. If the spin splitting is not zero but is still small compared to the cyclotron splitting ($\Delta_1/\Delta_2 \ll 1$, where $\Delta_1$ is the splitting between "spin-up" and "spin-down" sublevels of the 0th Landau level and $\Delta_2$ is the splitting between the "spin-down" sublevel of the 0th Landau level and the "spin-up" sublevel of the 1st Landau level), the phase diagram becomes the one suggested by Fogler and Shklovskii (Fig. 1 (b)). The only allowed transition for increasing $B$ is still $0 \rightarrow 2$ transition. This situation was experimentally observed in Refs. [2-4] and is schematically shown by the dashed line in Fig. 1 (b). If the disorder exceeds a certain critical value, as in these experiments, the spin splitting is not observed. However, it is clear that if the ratio $\Delta_1/\Delta_2$ is further increased, the spin splitting will survive for stronger disorder; this corresponds to the triple point (solid circle) in Fig. 1 (b) moving up the curve to higher values of disorder. In the (unrealistic but instructive) limit of $\Delta_1/\Delta_2 = 1$, all energy levels would be equally spaced, and one obtains the phase diagram equivalent to that for spinless electrons proposed by Kivelson, Lee, and Zhang [6] and shown in Fig. 1 (d). The phase diagram of Fig. 1 (c), equivalent to that reported in Ref. [1], is an intermediate case of $\Delta_1/\Delta_2 < 1$. The dashed line in Fig. 1 (c) schematically shows the transition observed in Ref. [1].

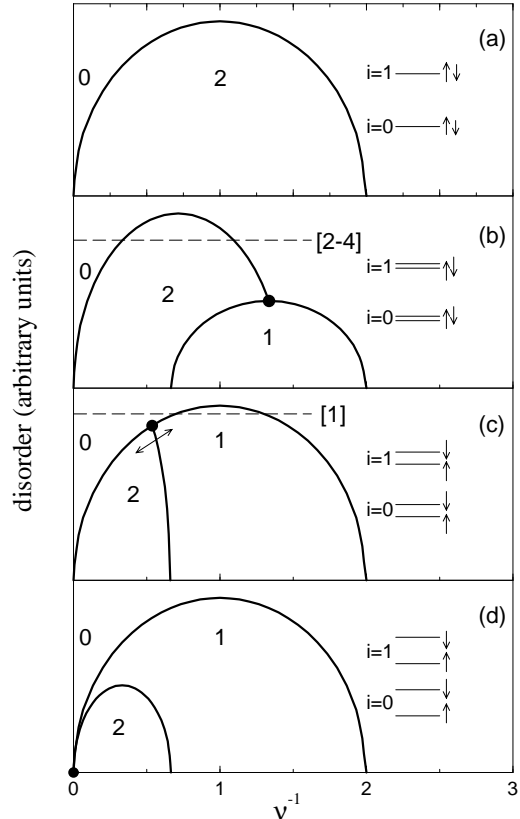

FIG. 1. Schematic "disorder vs inverse filling factor" phase diagrams for different ratios between spin and cyclotron splittings: (a) $\Delta_1/\Delta_2 = 0$, (b) $\Delta_1/\Delta_2 \ll 1$, (c) $\Delta_1/\Delta_2 < 1$, and (d) $\Delta_1/\Delta_2 = 1$. Numbers 0, 1, 2 give $\sigma_{xy}$ in units of $e^2/h$. Dashed lines show schematically insulator—QHE transitions consistent with the experimental references with which they are labeled. To the right of the diagrams, the corresponding level structures are shown with arrows representing spin directions; Landau levels are identified by $i = 0$ and 1.

The splitting at $\nu = 1$ is an exchange-enhanced spin splitting which in first-order perturbation theory in the case of single-particle excitations (no skyrmions) is equal to [7]

$$\Delta_1|_{\nu=1} \sim g\mu_B B + (\pi/2)^{1/2} \frac{e^2}{\epsilon l_B} \qquad (1)$$

(here $g$ is the Lande $g$-factor, $\mu_B$ is the Bohr magneton, $e$ is the electron charge, $\epsilon$ is the dielectric constant, and $l_B$ is the magnetic length). The first term in this equation is the Zeeman gap and is usually much less than the second one which represents the many-body enhancement of the spin gap. Therefore, $\Delta_1$ is roughly proportional to $l_B^{-1} \propto B^{1/2}$. The splitting $\Delta_2$ is the cyclotron splitting minus the spin splitting, and since the latter is usually much smaller than the first, $\Delta_2$ is of the order of

$$\Delta_2|_{\nu=1} \sim \hbar\omega_c \propto B \qquad (2)$$

($\omega_c$ is the cyclotron frequency). Hence, the ratio

$$\Delta_1/\Delta_2|_{\nu=1} \sim B^{-1/2} \qquad (3)$$

increases as $B$ is decreased. Of course, in a real 2D system splittings between levels are less than those given by Eqs. (1) and (2) because of nonzero level widths and other corrections, but qualitative trend will be the same.

For a given filling factor, smaller electron densities require proportionally smaller magnetic fields, which in turn yield larger values of $\Delta_1/\Delta_2$. The samples used by Shahar, Tsui, and Cunningham [1] have electron densities that are more that the order of magnitude lower than those used in, e.g., Ref. [3]. Therefore, the ratio $\Delta_1/\Delta_2$ for their samples exceeds the corresponding ratio for samples of Ref. [3] by the square root of 10, or more than a factor of 3. This explains why the phase diagram of Ref. [1] is qualitatively different.

In the above consideration I did not consider skyrmions, which according to Ref. [7] are the only relevant excitations at $\nu = 1$ for small g-factor. However, the energy of skyrmions is also of the order of $e^2/\epsilon l_B \propto B^{1/2}$. Therefore the above speculations remain in force. A more important problem is that the ratio $\Delta_1/\Delta_2$, and hence the phase diagram, depends strongly on the filling factor. The exchange enhancement of the spin splitting occurs only when the Fermi energy enters the gap between two spin-split levels, i.e., at $\nu < 2$; therefore, in a GaAs/AlGaAs heterostructure, where the bare, unenhanced spin splitting is much less that the cyclotron splitting ($g\mu_B B \ll \hbar\omega_c$), the triple point in Fig. 1 cannot move to the left of $\nu^{-1} = 0.5$. In the experiments reported in Ref. [1], the triple point corresponds to $\nu^{-1} \approx 0.7$ which does not contradict the above consideration.

A question may arise why the phase diagram of Ref. [1] does not have a "camel-back" structure reported in Ref. [8] for 2D system based on silicon. A possible reason is that the local maxima in the phase diagram of Ref. [8] were observed at filling factors 1, 2, and 6, corresponding, in the case of silicon, to valley ($\nu = 1$) and spin ($\nu = 2$ and 6) splittings which are known to be subject to strong many-body enhancement. A mechanism how this many-body enhancement may lead to the maxima in the phase diagram ("camel-back" structure) was suggested in Ref. [8]. On the other hand, in a GaAs/AlGaAs heterostructure, the splitting at $\nu = 2$ is a *cyclotron* splitting which has small many-body enhancement and therefore there is no maximum at $\nu = 2$ in this system.

I am grateful to H. Fertig, M. Sarachik, and D. Simonian for useful discussions. This work was supported by the US Department of Energy under Grant No. DE-FG02-84ER45153.